\documentclass[12pt]{article}
\usepackage{refmerge}
\usepackage{epsfig}
\begin{document}
\begin{center}
{\bf{ \boldmath
FIRST OBSERVATION OF THE $\phi\to\pi^+\pi^-\gamma$ DECAY 
}}
\end{center}
\begin{center}
R.R.Akhmetshin, E.V.Anashkin, M.Arpagaus, V.M.Aulchenko,
V.S.Banzarov, L.M.Barkov, N.S.Bashtovoy, A.E.Bondar, D.V.Bondarev,
A.V.Bragin,
D.V.Chernyak, A.S.Dvoretsky,
S.I.Eidelman, G.V.Fedotovich, N.I.Gabyshev,
A.A.Grebeniuk, D.N.Grigoriev, P.M.Ivanov, S.V.Karpov, V.F.Kazanin,
B.I.Khazin, I.A.Koop, P.P.Krokovny,
L.M.Kurdadze, A.S.Kuzmin,
I.B.Logashenko, P.A.Lukin, A.P.Lysenko, K.Yu.Mikhailov, 
I.N.Nesterenko, V.S.Okhapkin, E.A.Perevedentsev, E.A.Panich, A.S.Popov, 
T.A.Purlatz, N.I.Root, A.A.Ruban, N.M.Ryskulov,
A.G.Shamov, Yu.M.Shatunov, B.A.Shwartz, A.L.Sibidanov, V.A.Sidorov,
A.N.Skrinsky, V.P.Smakhtin, I.G.Snopkov, E.P.Solodov, 
P.Yu.Stepanov, A.I.Sukhanov, 
V.M.Titov, Yu.V.Yudin, S.G.Zverev
\\
  Budker Institute of Nuclear Physics, Novosibirsk, 630090, Russia
\end{center}
\begin{center}
                J.A.Thompson
\\
                University of Pittsburgh, Pittsburgh, PA 15260, USA
\end{center}
\vspace{0.7cm}
\begin{abstract}
\hspace*{\parindent}
Radiative decays of the $\phi$ meson have been studied
using a data sample of about 20 million $\phi$ decays
collected by the CMD-2 detector at VEPP-2M collider in Novosibirsk.
From selected
$e^+e^-\to\pi^{+}\pi^{-}\gamma$  
events the $\phi\to\pi^{+}\pi^{-}\gamma$ decay has been observed 
for the first time.

 Under the assumption that the intermediate  
$f_{0}(980)\gamma$ state dominates 
in the $\phi \to \pi^+\pi^-\gamma$ decay, 
the corresponding branching ratio is
 $Br(\phi\rightarrow f_{0}(980)\gamma)=(1.93\pm 0.46\pm 0.50)\times10^{-4}$.

Selected  $e^+e^-\rightarrow\mu^+\mu^-\gamma$ events were used to obtain
 $Br(\phi\rightarrow\mu^+\mu^-\gamma) =(1.43\pm 0.45\pm 0.14)\times10^{-5}$ 
for $E_{\gamma}>20$ MeV.

Using the same data sample, upper limits at 90\% CL have been obtained  
for the C-violating decay of the $\phi$:
 $Br(\phi\rightarrow\rho\gamma) < 1.2\times10^{-5}$;
and for the P- and CP-violating 
decay of the $\eta$:

 $Br(\eta\rightarrow\pi^+\pi^-) < 3.3\times10^{-4}$. 
\end{abstract}
\baselineskip=17pt
\section*{ \boldmath Introduction}
\hspace*{\parindent}
 The identification of scalar mesons and
particularly the determination of the lightest scalar $q\bar{q}$ nonet 
is of extreme importance both for quark systematics and searches for 
non-$q\bar{q}$ objects like glueballs and multiquark states (see the
discussion of this problem in the minireview on scalar mesons by
S.Spanier and N.T\"ornqvist, p.390 of Ref. \cite{pdg}).      
It is generally accepted that the probabilities of the electric dipole 
radiative transitions of the $\phi$ meson are crucial for the 
clarification of the nature of $f_0(980)$ and $a_0(980)$ 
mesons \cite{AchIvan,close}. A search for 
these decays has been earlier performed by the ND~\cite{dol91} and
CMD-2~\cite{CMDppg} groups. A number of new
results has been recently published  
by  SND~\cite{SNDa0g,SNDppg} and
CMD-2~\cite{ICHEP98,CMD9911} 
collaborations at the 
$e^+e^-$ collider VEPP-2M~\cite{vepp2m} where the
radiative
decays $\phi\to\pi^{0}\pi^{0}\gamma$, $\phi\to\eta\pi^{0}\gamma$ and
$\phi\to\pi^+\pi^-\gamma$ have been observed for the first time. 

The mode with two charged pions has very large background because of
the radiative processes 
$e^+e^-\to\pi^+\pi^-\gamma$ where a photon comes from initial 
electrons or from final pions. Therefore a signal from the 
$f_{0}(980)\gamma$ final state can be  seen  
as an interference structure at the energy  
 $E_{\gamma} = \frac {m_{\phi}^2-m_{f_{0}^2}} {2m_{\phi}} \approx $  
40 MeV in the photon spectrum. The shape of this structure 
depends on the $f_{0}(980)$ mass  and width. 
 In our first paper ~\cite{CMDppg} a search for the 
$\phi\to\pi^+\pi^-\gamma$ decay led to the upper limit of
$3\times 10^{-5}$ that looked at first puzzlingly low compared to the decay
$\phi\to\pi^{0}\pi^{0}\gamma$ observed with the branching ratio
of about $1\times 10^{-4}$~\cite{SNDppg}. 
As was shown in  ~\cite{ICHEP98,CMDppg,acha97}, 
in case of the $\phi\to f_{0}(980)\gamma$ intermediate state it could be
explained by the destructive interference between bremsstrahlung processes
and the $\phi$ decay.

In this paper results of the study of the
$e^+ e^- \to\pi^+\pi^-\gamma$ process with the CMD-2 detector~\cite{CMD285} 
are presented.
In total, the 14.2 $pb^{-1}$ of data have been
collected  since 1993 at
14 energy points around the $\phi$ mass.
For the analysis of the $\pi^+\pi^-\gamma$ decay mode
13.1 $pb^{-1}$ corresponding to  $19.7 \times 10^{6}$ $\phi$ decays
were used.
Seven scans of this energy
region were performed allowing control of systematic errors caused by 
possible detector instability. The results obtained 
from individual scans were found to be consistent.

\section*{ \boldmath Selection of $\pi^+\pi^-\gamma$ Events} 
\hspace*{\parindent}
Event candidates were selected by requiring  only  two
minimum ionizing tracks in the drift chamber (DC)
and one or two photons with energy greater than 20 MeV in
the CsI calorimeter.
The following selection criteria were used: 
\begin{enumerate}
\item{
The average momentum of two charged particles is higher than 240 MeV/c 
to remove the background from $K_{S} \rightarrow \pi^{+}\pi^{-}$ decays.
}
\item{
Detected tracks have a polar angle between
1.05 and 2.1 radians so that they enter the inner muon system.
}
\item{
The sum of the energy depositions of two clusters 
associated with two tracks is less than 450 MeV 
to remove Bhabha events.
}
\item{
The radial distance of the closest approach of each track to the 
beam axis is less than 0.3 cm.
}
\item{
The Z-coordinate of the vertex (along the beam) is 
within 10 cm from the detector center. This cut reduces cosmic ray 
background by a factor of two.
}
\item{
Detected photons have a polar angle between
0.85 and 2.25 radians so that they enter
the "good" region in the CsI barrel calorimeter.
This requirement suppressed the background from the photons emitted by 
initial electrons. 
}
\item{
Events with an invariant mass of two photons close to the
$\pi^0$ mass ($|m_{\gamma\gamma}-m_{\pi^0}|<40$ MeV) were removed.
}
\end{enumerate}    

After the above cuts the main  background for the $\pi^{+}\pi^{-}\gamma$ 
final state comes from:
a) the radiative process $e^+ e^- \rightarrow\mu^+\mu^-\gamma$,
b)  the decay $\phi \rightarrow \pi^{+}\pi^{-}\pi^{0}$ when one of the photons
from the $\pi^0$ escapes detection and
c) collinear events $e^{+}e^{-} \rightarrow \mu^{+}\mu^{-},\pi^{+}\pi^{-}$
in which secondary decays
and interactions of muons or pions with the detector material
produce a background cluster mimicking a photon.

%
\begin{figure}[tbh]
\epsfig{figure= 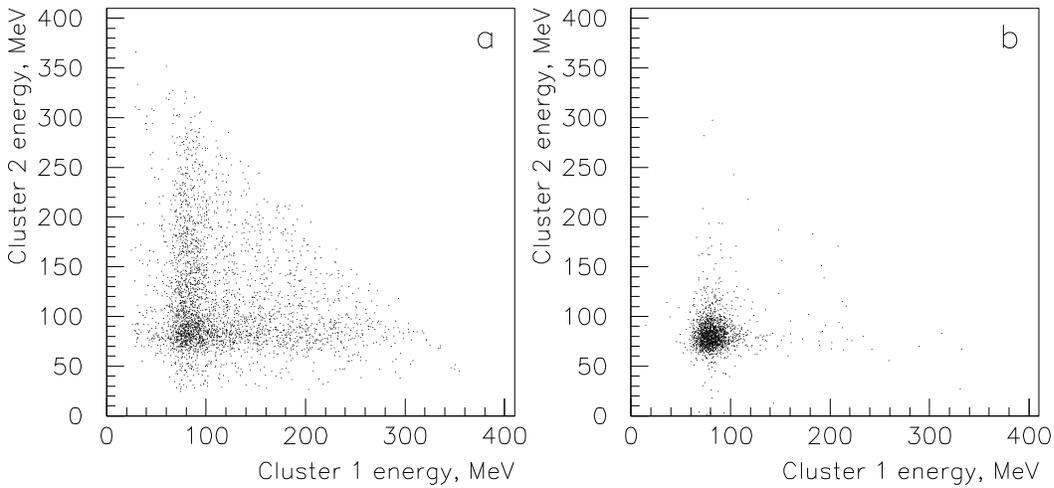,width=1.0\textwidth}
\caption{Calorimeter response for events with one or no hits in 
the muon system (a) and for events selected as muons (b).}
\label{ec1-ec2}
\end{figure}
%

%
\section*{ \boldmath Selection of $\mu^+\mu^-\gamma$ Events}
\hspace*{\parindent}
The inner muon  system was used to separate muons from pions.
The muon  system uses  streamer tubes grouped in two layers (inner and
outer) with a 15 cm magnet yoke serving as an absorber and has 1-3 cm spatial 
resolution.
The requirement of hits 
in the inner muon system for both charged particles selects muon events,
together with some pion events in which both pions pass 
the calorimeter without nuclear interaction.

Separation of pion and muon events in the CsI calorimeter
is illustrated in Fig.~\ref{ec1-ec2}, where
scatter plots of the energy deposition of one
track vs. that of the other one
are presented for events with one or no hits in the muon system (a) and
selected as muons (b). Energy depositions are
corrected for the incident angle.  Pions can have nuclear interactions and 
in some cases leave more energy, while muons mostly exhibit dE/dx losses
only.

%
\begin {figure} [tbh]
\hspace{0.5cm}
\psfig{figure= 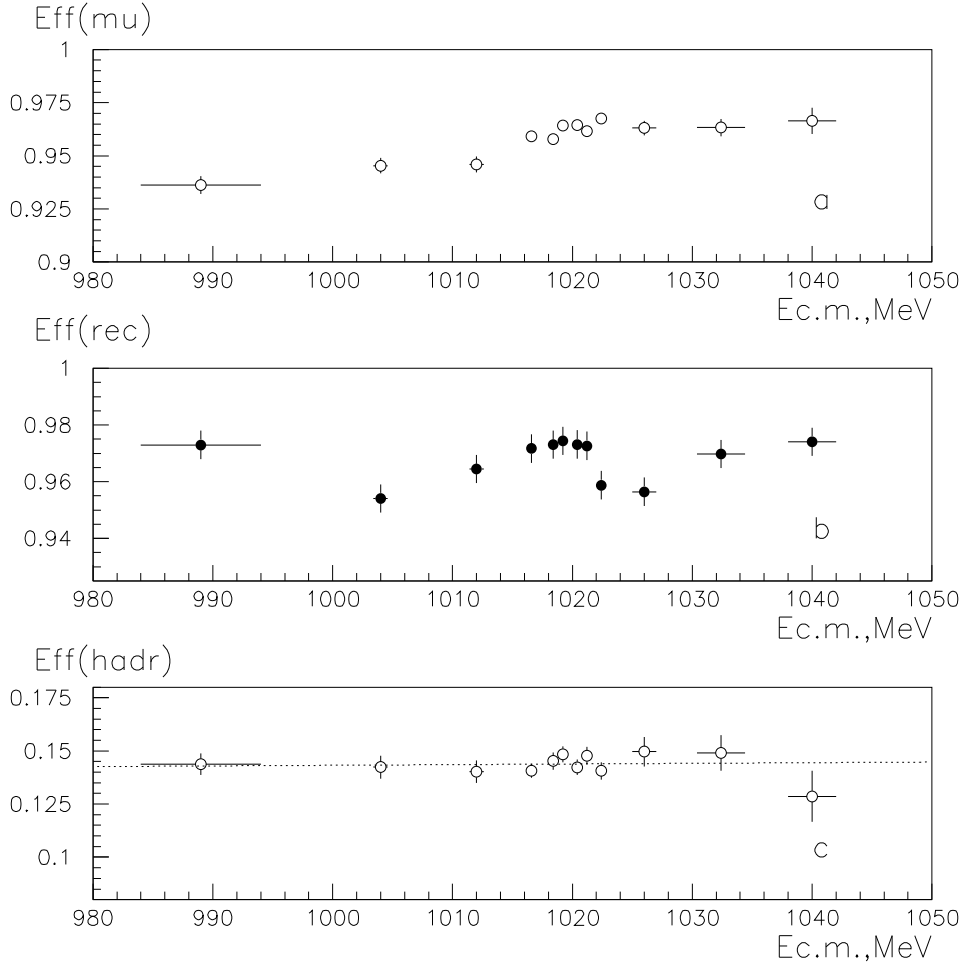,width=0.9\textwidth,  height=0.8\textwidth}
\caption{
a.Muon system efficiency vs. energy.
b.Reconstruction efficiency for DC.
c.Probability for two pions to be selected as muons. 
} 
\label{pimu_eff} 
\end{figure}
To select a cleaner sample of  muon events,
in  addition to the information from the muon system 
both tracks were required to show only minimum ionizing energy deposition
in the calorimeter (60-130 MeV).
All the remaining events were considered as pion candidates.
In this raw separation
the pion sample $N_{\pi\pi\gamma}$ contains muons because of 
some inefficiency of the muon system
while the muon sample $N_{\mu\mu\gamma}$ contains pions since a  
pion can reach the muon system without nuclear interaction. 
 The observed numbers of events $N_{\pi\pi\gamma}$ and
$N_{\mu\mu\gamma}$ are related to the true numbers  
$N^0_{\pi\pi\gamma}$ and $N^0_{\mu\mu\gamma}$ as:\\
$N_{\pi\pi\gamma} = (1- \epsilon_{\mu})\cdot N^0_{\mu\mu\gamma} +(1- \epsilon_{\pi})\cdot N^0_{\pi\pi\gamma}$,\\  
$N_{\mu\mu\gamma} = \epsilon_{\mu}\cdot N^0_{\mu\mu\gamma} + 
\epsilon_{\pi}\cdot N^0_{\pi\pi\gamma}$,\\ 
\noindent
where $\epsilon_{\mu}$ and $\epsilon_{\pi}$ are the muon system 
efficiency and probability for a pion pair to be selected as muons 
respectively.

The magnitudes of  $\epsilon_{\mu}$ and $\epsilon_{\pi}$ were 
determined by studying correlations 
between the energy deposition in the CsI calorimeter and response of the 
muon system to collinear $\pi^+\pi^-$ and $\mu^+\mu^-$ events. 
The results of this study are shown in Fig.~\ref{pimu_eff} for one of the 
experimental scans.
Figure~\ref{pimu_eff}a presents the muon system efficiency for different
energy points. Only statistical errors are shown. 
The probability for two pions to be selected as muons is presented in
Fig.~\ref{pimu_eff}c and is independent of the possible
instability of the detection system.

The number of pions and muons has been corrected for the DC 
reconstruction efficiency shown in Fig.~\ref{pimu_eff}b. This
efficiency  was determined from
collinear Bhabha events as described in \cite{CMD9911}.

Our study of the efficiencies gives a correlated 3\% systematic 
uncertainty in the final number of $\pi^+\pi^-$ and $\mu^+\mu^-$ events.

%
\begin{figure}[tbh]
\begin{center}
\epsfig{figure= 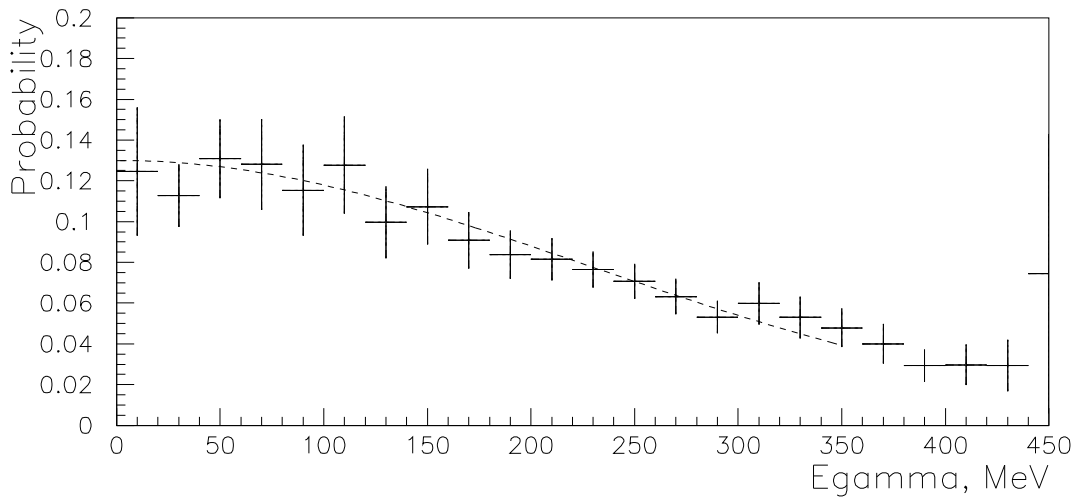,width=1.0\textwidth}
\caption{
Simulated probability for pions to be selected as muons vs. photon 
energy.
}
\label{pi_loss}
\end{center}
\end{figure}
%
For events with photons $\epsilon_{\mu}$ and $\epsilon_{\pi}$ depend on
photon energy. This dependence   
was studied using
simulated $\pi^+\pi^-\gamma$ events. The probability of two pions to be 
selected as muons varies with photon energy as presented in Fig.~\ref{pi_loss} 
and was used to correct the final number of events with photons.
At low photon energy this probability is consistent with that obtained
from collinear events (see Fig.~\ref{pimu_eff}c).
\section*{ \boldmath Constrained Fit}
\hspace*{\parindent}
To reduce the background from collinear events as well as that from the
three pion $\phi$ decay a constrained fit was used requiring total 
energy-momentum conservation for a three body decay. 
About 20\% of the selected events had an additional photon. In this case
the constrained fit was applied to both possible combinations and 
that with a minimum $\chi^2$ was chosen.

An additional cut was applied to the photon direction: an azimuthal angle
should be more than 0.25 radians from the charged track direction. 
This cut removed the remaining collinear events
with a background photon which survived after the constrained fit. 

\begin{figure}[tbh]
\begin{center}
\epsfig{figure= 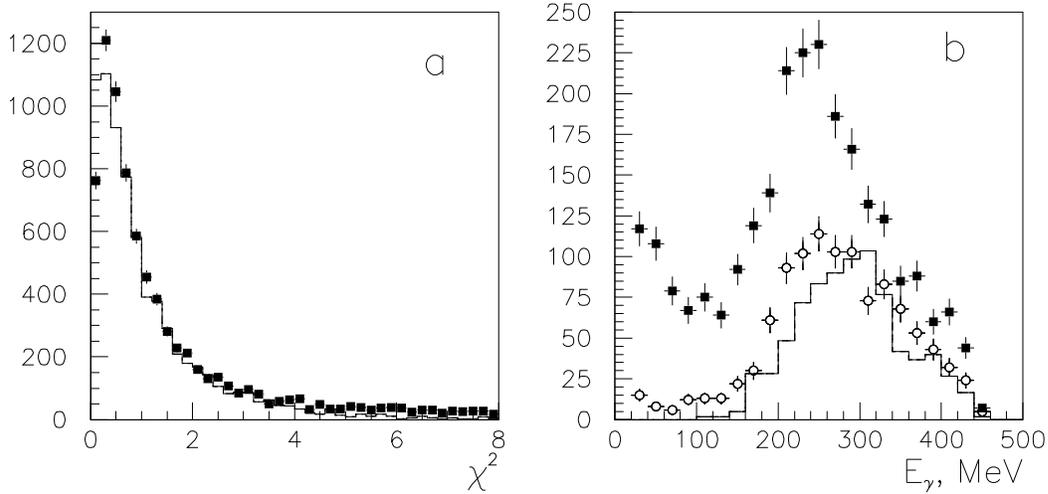,width=1.0\textwidth}
\caption{
a. The $\chi^2$/d.f. distribution for events selected as muons. The histogram
is simulation.
b. Photon spectrum for events selected as pions for  $\chi^2$/d.f.$<$3
(dark points) and for 3 $<~\chi^2$/d.f.$~<$ 6 (open points). The histogram
is simulation of  background from $\pi^+\pi^-\pi^{0}$ events.
}
\label{xi2_pmg}
\end{center}
\end{figure}

The $\chi^2$/d.f. distribution for events selected as muons had very
small background and was found to be in good agreement with simulation
as shown in Fig.~\ref{xi2_pmg}a.
Motivated by simulation,
a cut that $\chi^2$/d.f. be less than 3 was imposed for pion and 
muon events selecting 95\% of signal events.

After the above cut the pion sample still 
contained some background mainly from three pion $\phi$ decays.
Such three pion background appears when one of the photons 
from the $\pi^{0}$ has energy below 20 MeV and is not detected
so that the event looks like a three body decay with the remaining photon
energy higher than 150 MeV.
The $\chi^2$/d.f distribution for these events was flat and 
events with 3 $<~\chi^2$/d.f.$~<$ 6,
 shown in Fig.~\ref{xi2_pmg}b by open points,
 were used to estimate the
background spectrum.
At each energy
the background spectrum was subtracted from that for the signal candidates.

\begin {figure} [tbh]
\vspace{-1.0cm} 
\epsfig{figure= 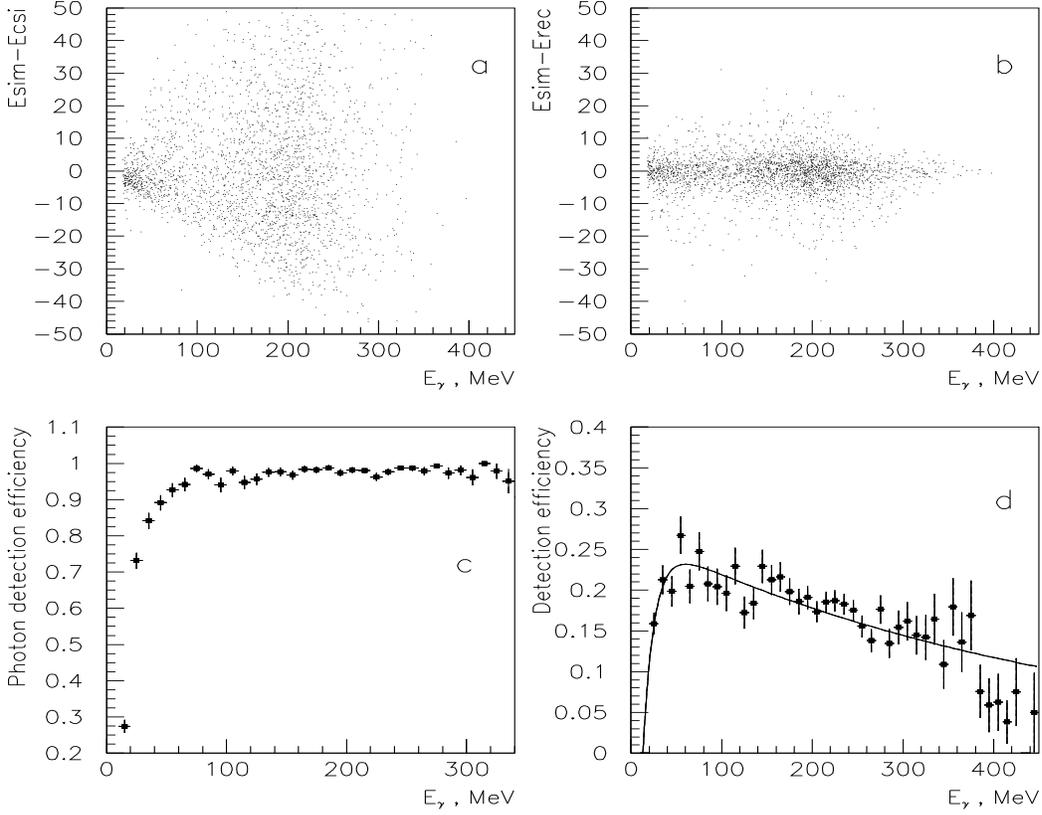,width=1.0\textwidth,  height=0.8\textwidth} 
\caption {Study of simulated  $\pi^+\pi^-\gamma$ events.
a. Difference between the photon energy measured in the calorimeter and 
the initially simulated one vs. photon energy.
b. Difference between the kinematically reconstructed photon energy and 
the initially simulated one vs. photon energy. 
c. Simulated photon detection efficiency in the CsI calorimeter.
d. Overall detector efficiency vs. photon energy.
} 
\label {gam_eff} 
\end {figure}

From a constrained fit one can
obtain an improved estimate for the photon energy. This effect was studied by
using simulation and results are presented in Fig.~\ref{gam_eff}. 
Simulation shows that after the constrained fit photons have energy
resolution about 5 MeV in the whole 
energy range instead of $\sigma_{E_{\gamma}}=8\%\times E_{\gamma}$ 
CsI resolution as shown in Figs.~\ref{gam_eff}a,b. 
Figure~\ref{gam_eff}c demonstrates the simulated photon detection efficiency in
the CsI calorimeter vs. photon energy. The overall detector efficiency
vs. photon energy  
is shown in Fig.~\ref{gam_eff}d for
$\pi^+\pi^-\gamma$ events.

To extract the resonant contribution associated with the
$\phi$, two data sets were used. 
The first set (``$\phi$'' region)   for 
E$_{c.m.}$ from 1016.0 to 1023.2 MeV had the integrated luminosity of
9.24~$pb^{-1}$ while the second one ("off-$\phi$" region) taken at 
E$_{c.m.}$ = 996-1013 and 1026-1060 MeV with the  
integrated luminosity of 3.89~$pb^{-1}$ containing less than 3\%  
$\phi$ decays was used for a background estimate.

\begin {figure} [tbh] 
\vspace{-0.5cm}
\epsfig{figure= 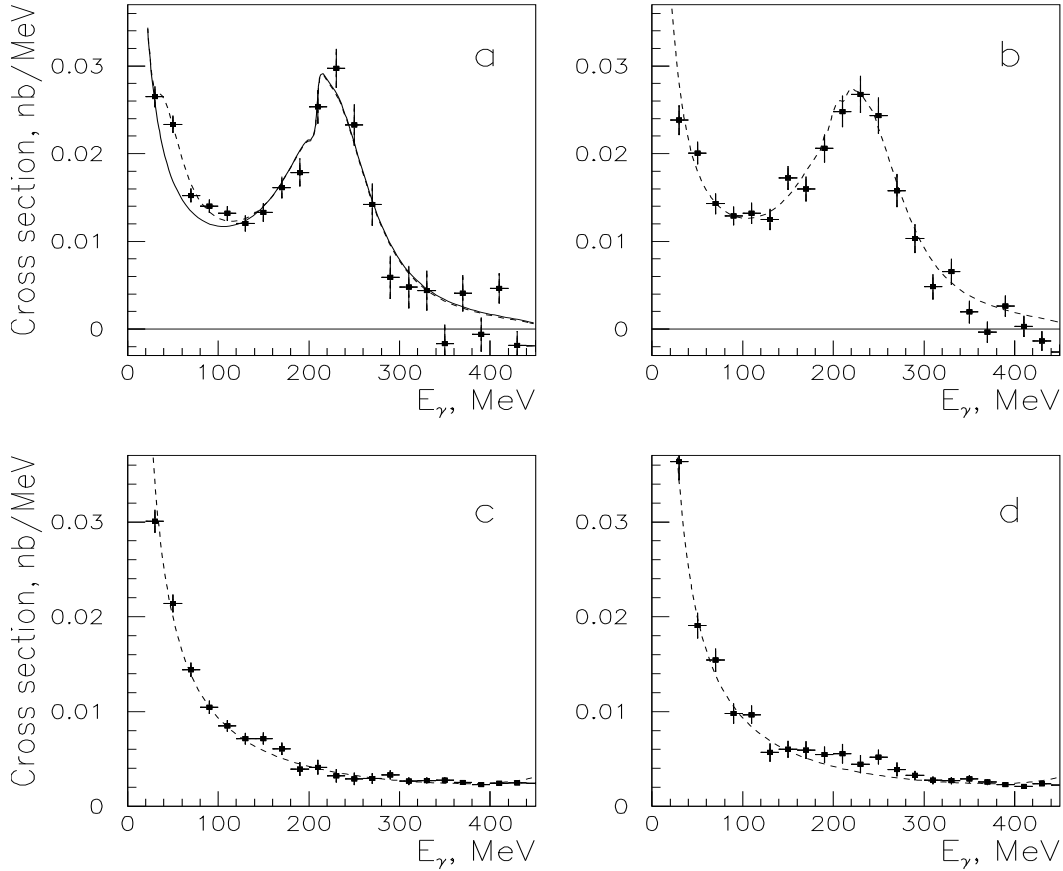,width=1.0\textwidth, height=0.8\textwidth} 
\vspace{-1.0cm}
\caption {
a. Photon spectrum for 
 $\pi^ + \pi^-\gamma$ events at the "$\phi$" region; Solid line is pure
bremsstrahlung. Dashed lines include a possible $\phi$ signal according
to ~\cite{acha97}.
b. Photon spectrum for 
 $\pi^ + \pi^-\gamma$ events at the "off-$\phi$" region;
c,d. The same spectra for
 $\mu^ + \mu^-\gamma$ events.
} 
\label {gam_sum} 
\end {figure}

Figure \ref{gam_sum} presents photon spectra obtained
after background subtraction and corrections for the detector efficiency
at the "$\phi$" (a,c) and "off-$\phi$" (b,d) regions for pions and
muons respectively. The solid line corresponds to the theoretical
calculation~\cite{acha97} taking into account the integrated luminosity 
at each energy point and $\rho - \omega$ mixing for the bremsstrahlung 
process.
A peak at 220 MeV is due to the radiation by initial electrons 
resulting in the process 
$e^+e^-\rightarrow\rho\gamma$,  $\rho \to \pi^+\pi^-$.
The photon energy range 20 to 120 MeV has minimum background 
and some excess of events 
over the expected bremsstrahlung spectrum (the solid line in 
Fig.~\ref{gam_sum}a) can be
seen in the $\pi^ + \pi^-\gamma$ sample at the "$\phi$" region.
In total, 30175 $\pi^ + \pi^-\gamma$ events and 27188 $\mu^+\mu^-\gamma$
events have been selected in this energy range. 
\section*{Cross Section Study}
\hspace*{\parindent}
The cross section for each energy point was calculated as $\sigma$ = 
$N_{ev}$/(L$\cdot\epsilon$).
$N_{ev}$ is the number of selected events 
with photons in the energy range from 20 to 120 MeV.
The integrated luminosity L for each energy point was determined from Bhabha
events with about 2\% systematic accuracy~\cite{CMD9911}.
The detection efficiency $\epsilon$ was obtained by simulation 
and the approximation shown by the solid line in Fig.~\ref{gam_eff}d 
was used to correct photon spectra. 

\begin {figure} [tbh]
\vspace{-1.0cm} 
\epsfig{figure= 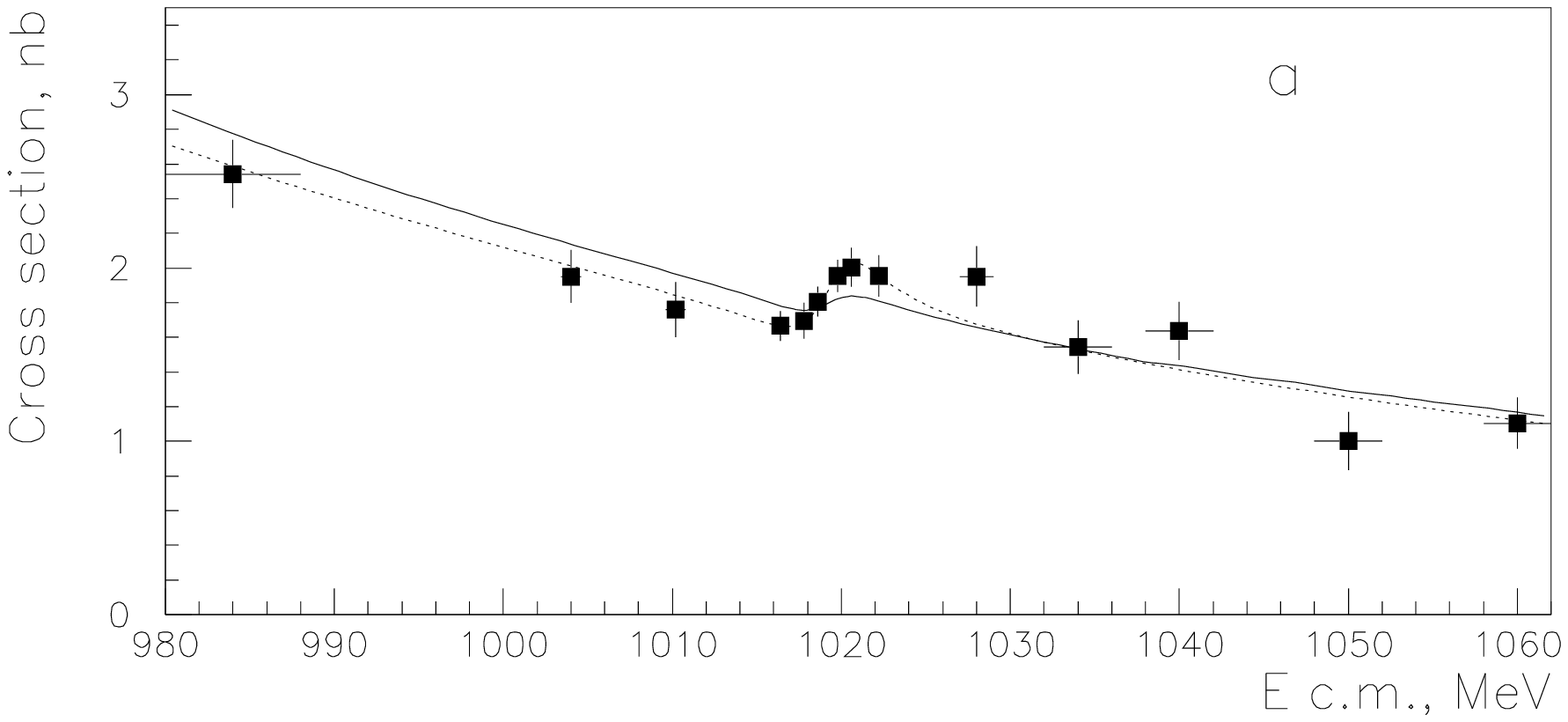,width=1.0\textwidth}
\epsfig{figure= 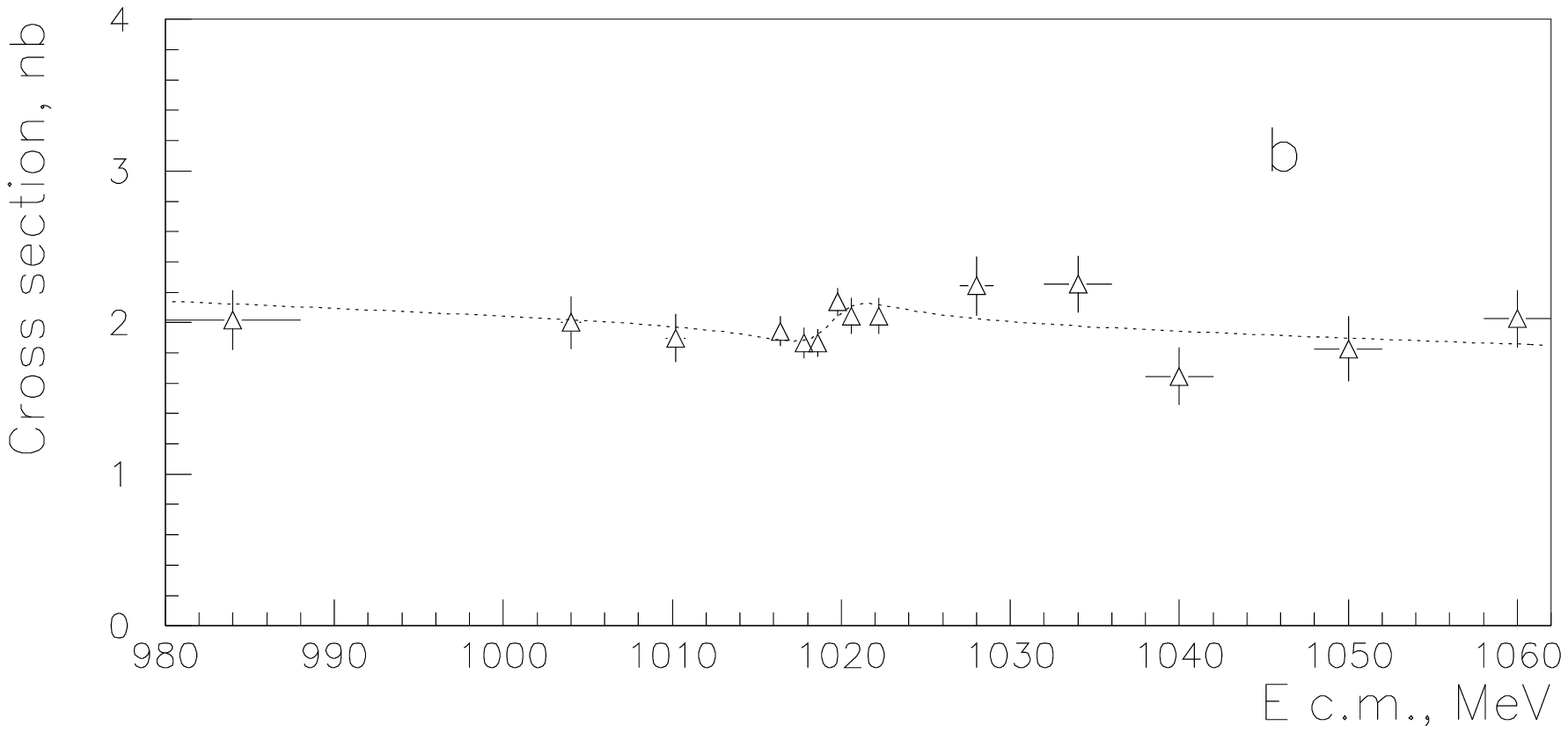,width=1.0\textwidth}
\caption{
a. Cross section for 
$e^{+}e^{-}\rightarrow\pi^{+}\pi^{-}\gamma$. Lines are theoretical
predictions in case of no direct $\phi$ decay (solid line) and best fit
(dotted line).
b. Cross section for $e^{+}e^{-}\rightarrow\mu^{+}\mu^{-}\gamma$
with the theoretical prediction.
}
\label{fnot-gamma}
\end{figure}

The resulting cross sections of the processes
$e^+e^-\rightarrow\pi^+\pi^-\gamma$ and
$e^+e^-\rightarrow\mu^+\mu^-\gamma$  versus energy
are presented in Fig. \ref{fnot-gamma}a,b. Only statistical errors are
shown.
The systematic error in the experimental cross sections was estimated to
be about 5\% dominated by the uncertainty of the pion-muon separation 
efficiencies. 

The cross sections can be described 
according to the calculations performed
in ~\cite{AchIvan,acha97}.
It has been shown there that the amplitudes describing initial bremsstrahlung
differ from those for the
final bremsstrahlung and the $\phi$ meson, so that
the differential cross section for $\pi^+\pi^-\gamma$
events is: \\

\noindent
$
\frac{d\sigma(s,E_{\gamma})}{dE_{\gamma}}\approx 
|A_{br}^{in}(s,E_{\gamma})|^2 +|A_{br}^{f}(s,E_{\gamma}) +  
A_{\phi}(s,E_{\gamma}) \pm
e^{i\Psi}\cdot A_{f_{0}}(s,E_{\gamma})|^2
$.\\

\noindent
Here $A_{br}^{in}(s,E_{\gamma})$ and $A_{br}^{in}(s,E_{\gamma})$
are the amplitudes describing initial and
final bremsstrahlung processes, and
the amplitude $A_{\phi}(s,E_{\gamma})$ introduces the influence of the
 $\phi$ upon the  photon propagator (a vacuum polarization term).
 This 
contribution gives rise to an interference pattern  in the cross section 
at the $\phi$ mass and can be referred to as an "electromagnetic" decay
$\phi\to\gamma\to\pi^+\pi^-\gamma$ or
$\phi\to\gamma\to\mu^+\mu^-\gamma$.
The amplitude of the interference is determined by the  
$\phi$ meson leptonic coupling constant. 

 The amplitude $A_{f_{0}}(s,E_{\gamma})$ represents the 
possible  $\phi$ decay into the $\pi^+\pi^-\gamma$ final state
via $f_0\gamma$. For  $\mu^+\mu^-\gamma$ events  
$A_{f_{0}}(s,E_{\gamma})$ = 0.
The model considers the $f_{0}$ meson as a two- or
four-quark state or $K\bar K$ molecule depending on the values of the coupling
constants g$^{2}_{K\bar{K}}/4\pi$ and  g$^{2}_{\pi\pi}/4\pi$. 
 
It also includes the interference of the $\phi$
amplitude  with the final
bremsstrahlung  amplitude $A_{br}^{f}(s,E_{\gamma})$
as well as the correction for the   
 $\pi^+ \pi^-$ scattering in the final state~\cite{acha98}. The
latter effect gives a contribution of
about 15\% to the branching ratio, but
within our accuracy can be introduced as
an additional phase
shift $\Psi$ between the amplitudes of the $\phi$  and  bremsstrahlung process.
This shift was predicted to be 1.2-1.4 radians, but the sign of the term
 was unknown and was  determined from the fit.
In general, all amplitudes have different dependence on s and $E_{\gamma}$. 

The initial bremsstrahlung process presented by the amplitude 
$A_{br}^{in}(s,E_{\gamma})$
is suppressed by 
selecting photons transverse to the beam 
direction, but it still accounts for about 2/3~\cite{acha97} of the observed  
$e^+e^-\to\pi^+\pi^-\gamma$ 
and one half of  the $e^+e^-\to\mu^+\mu^-\gamma$ cross section.

Integration over $E_{\gamma}$ from 20 to 120 MeV gives $\sigma(s)$ which can
be used for the cross section aproximation in Fig.~\ref{fnot-gamma}.

Assuming no hadronic 
$\phi\to\pi^{+}\pi^{-}\gamma$ decay   ($A_{f_{0}}(s,E_{\gamma})$=0)
a simple formula has been
used where the final radiation amplitude of the bremsstrahlung 
process interferes  with
the Breit-Wigner amplitude from the $\phi$ vacuum polarization:\\
 
$ 
\sigma(s) = \sigma^{in}_{br}(s) +
\sigma^{f}_{br}(s)\cdot 
| 1 - e^{i\psi}\cdot A_{\phi}^{(0)}
\frac{ m_{\phi}\Gamma_{\phi}} {\Delta_{\phi}} |^{2}$; \\
 
 $\Delta_{\phi} = s - m_{\phi}^{2} + i\sqrt{s} \Gamma_{\phi}(s)$.\\
  
Here $s=4E_{beam}^{2}$, $\Gamma_{\phi}$ and $m_{\phi}$
are $\phi$ meson parameters, $A_{\phi}^{(0)}$ is the $\phi$ decay 
amplitude in the peak and $\psi$ is the relative 
phase between the bremsstrahlung and $\phi$ decay amplitudes.

 This formula is valid under the natural assumption 
of the same photon spectrum for the final radiation amplitude of the 
bremsstrahlung process and 
the Breit-Wigner amplitude of the
"electromagnetic" $\phi$ decay.

The initial $\sigma^{in}_{br}(s)$ and final $\sigma^{f}_{br}(s)$  
bremsstrahlung cross sections have
different energy dependence for 20-120 MeV photons because of the pion
formfactor energy behaviour. 
According to \cite{acha97} the power functions
 $\sigma^{in}_{br}(s) =0.65\cdot\sigma_{0}\cdot(m_{\phi}/\sqrt{s})^{13}$ 
and
 $\sigma^{f}_{br}(s) =0.35\cdot\sigma_{0}\cdot(m_{\phi}/\sqrt{s})^{9}$ 
can describe the bremsstrahlung 
process $e^{+}e^{-}\to\pi^{+}\pi^{-}\gamma$ in our energy range.
The parameter $\sigma_{0}$ represents the sum of initial and final
bremsstrahlung  cross section at 
the $\phi$ mass.
The function $\sigma^{in}_{br}(s) = \sigma^{f}_{br}(s) = 0.5\cdot
\sigma_{0}(m_{\phi}^2/s)$  was used for muons.

The fit of experimental data with $\sigma_{0}$,
peak amplitude $A_{\phi}^{(0)}$ 
and $\psi$ as free parameters shows good agreement of obtained
cross sections with the theoretical calculations~\cite{acha97}:\\
$\sigma^{exp}_{0}/\sigma^{th}_{0} =1.02 \pm 0.02 \pm 0.05$,\\
$\sigma^{exp}_{0}/\sigma^{th}_{0} =0.97 \pm 0.02 \pm 0.05$ \\
for pions and muons respectively.
The above values also indicate that the $\mu~-~\pi$ separation 
uncertainty does not exceed the estimated systematic error.  

Using the obtained amplitude one can calculate the peak cross section  
 $\sigma (\phi\to\pi\pi\gamma) = 
|A_{\phi}^{(0)}|^2\cdot\sigma^{f}_{br}(m^2_{\phi})$ 
and the decay branching ratio can be calculated as
 $Br(\phi\to\pi\pi\gamma)=\sigma (\phi\to\pi\pi\gamma)/\sigma_{tot}^{\phi}$, 
where  $\sigma_{tot}^{\phi}$
is the total peak cross section of the $\phi$ resonance obtained from
the leptonic width~\cite{pdg}.
The following results have been obtained
after inserting corrections for angular acceptance equal to 0.68 and
0.42 for pions and muons respectively:

 $Br(\phi\to\pi^+\pi^-\gamma) = (0.41\pm0.12\pm0.04)\times10^{-4}$, 

 $Br(\phi\to\mu^+\mu^-\gamma) = (1.43\pm0.45\pm0.14)\times10^{-5}$,

 $\psi = (0.46 \pm 0.17) $ radians for pions and $\psi = (0.2\pm0.4)$
radians for muons. 

The obtained values should be compared to the theoretical calculations 
based on ``electromagnetic'' $\phi$ decays only~\cite{acha97}:

$Br(\phi\to\pi^+\pi^-\gamma$)=0.047$\times10^{-4}$,

$Br(\phi\to\mu^+\mu^-\gamma$)=1.15$\times10^{-5}$.\\
For muons the experimental value is in good agreement with theory, 
but for pions
the value of the measured branching ratio is 9 times larger 
than the theoretical expectation
and points to the presence of
the hadronic decay of $\phi$ to $\pi^+\pi^-\gamma$.
 The obtained branching ratios supersede our results in
\cite{CMDppg} based on the part of the total data sample.
The corresponding fit curves are shown in Fig.~\ref{fnot-gamma}. 
\section*{Studies of Photon Spectra}
\begin{figure}[tbh]
\epsfig{figure= 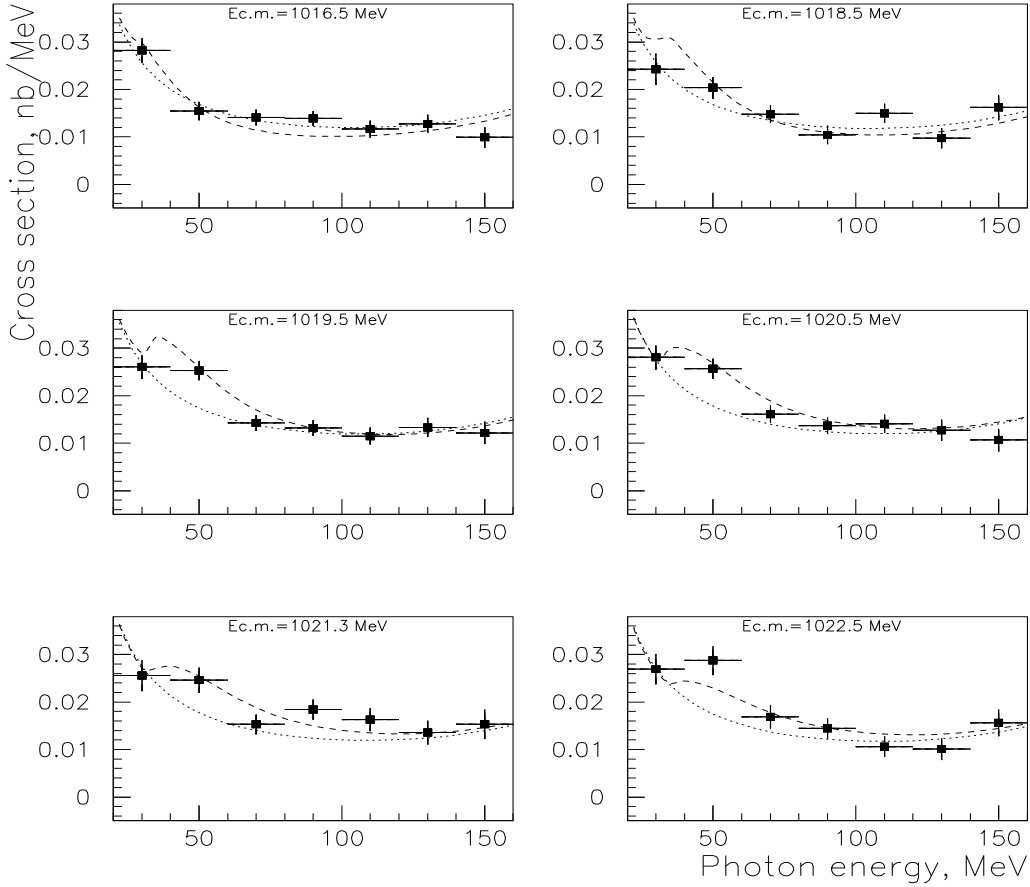,width=1.0\textwidth, height=0.8\textwidth}
\caption{
Photon spectra for six c.m.energy points in the "$\phi$" region.
Lines are theoretical predictions for the four quark model 
(dashed line) and pure bremsstrahlung spectra (dotted line).
} 
\label{f0-spectr}
\end{figure}

\hspace*{\parindent}
To search for the $\phi\to f_{0}(980)\gamma$ decay contribution
to the observed $\phi\to\pi^+\pi^-\gamma$ decay, photon energy spectra
were studied.
The photon spectra from the "$\phi$" region are  
shown in Fig.~\ref{f0-spectr} for six c.m.energy points 
and photons in the 20-160 MeV energy range. 
The spectra were corrected for all
experimental inefficiencies and normalized to the integrated luminosity.

 The signal from the decay of the $\phi\to f_{0}(980)\gamma$ is
seen as a structure in the photon spectra at 40-60 MeV.
Also shown by the dotted lines are the theoretical predictions  
for the expected bremsstrahlung spectra~\cite{acha97,acha98} including
vacuum polarization .

Because of the presence of the resonance in the photon spectra
the branching ratio obtained in the previous chapter for the
 $\phi\to\pi^+\pi^-\gamma$ decay can be used only as some indication 
to the existence of the $\phi$ hadronic decay into this mode.

It should be mentioned that
the excess of events over the bremsstrahlung spectra
cannot be used for the  branching ratio calculation.
As it was shown in ~\cite{ICHEP98,CMDppg,acha97}, the destructive 
interference with the bremsstrahlung process can reduce a visible
signal  in the selected energy range for photons.
To demonstrate the effect of interference the 
photon spectra in Fig.~\ref{f0-spectr}  
were fit as a group with the 
spectra 
calculated for each energy point using the differential cross section  
from ~\cite{AchIvan,acha97} (see previous section).

The number of events was not sufficient  to keep all model parameters free 
and therefore the model parameters g$^2_{K\bar{K}}/4\pi$ and 
g$^2_{\pi\pi}/4\pi$ were varied to keep the $f_{0}(980)$ width at about 
40 MeV \cite{pdg}. The fit had the following free parameters: the branching 
ratio,  $f_{0}(980)$ mass and phase shift $\Psi$.
 The data can be described
by the model
only for the destructive interference 
between the $f_{0}$ amplitude and the bremsstrahlung process.
The following results have been obtained:

 $Br(\phi\rightarrow f_{0}(980)\gamma)=(1.93\pm0.46\pm0.50)\times10^{-4}$;

 $m_{f_{0}}=976\pm5\pm6$ MeV;

 $\Psi = 1.55\pm0.22$ radians.\\
The results above include a systematic error from the 
weak dependence on the $f_{0}(980)$ width.
The relatively large value of the branching ratio obtained in the fit can
not be accounted for by the models assuming the normal $q\bar{q}$ structure
of  $f_{0}(980)$ \cite{somef0} and is 
predicted only in case of its four quark structure.
Note also that it
significantly differs from the branching ratio obtained from the
simple fit assuming the same photon spectra for all processes.
\section*{ \boldmath Search for $\phi\rightarrow\rho\gamma$ Decay}
\begin{figure}[tbh]
\epsfig{figure= 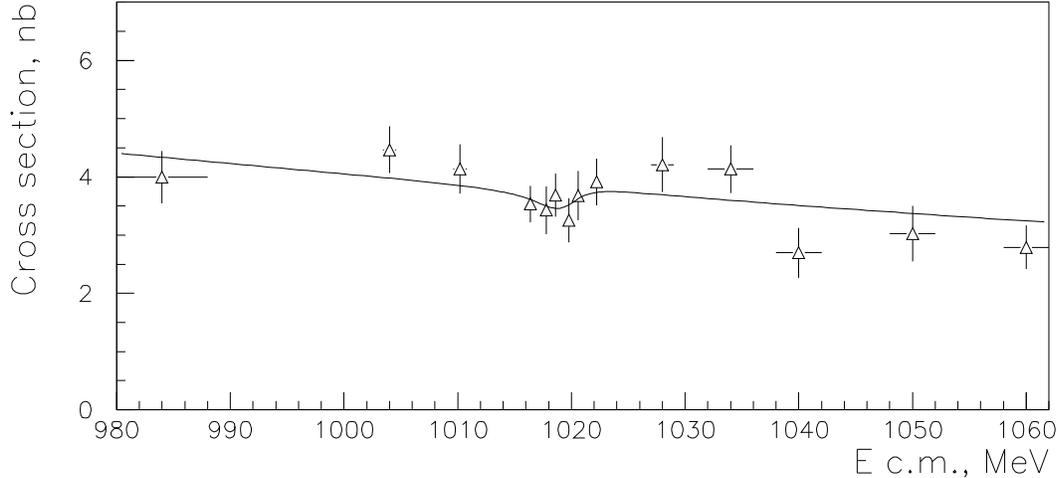, width=1.0\textwidth}
\caption{ $\phi\rightarrow\rho\gamma$ search.
The cross section for  $\pi^+\pi^-\gamma$ events with photons
in the 100-300 MeV range. The line is the best fit.
} 
\label{rho-sech}
\end{figure}

\hspace*{\parindent}
The selected $\pi^+\pi^-\gamma$ events with photon energies from
100 to 300 Mev (see Fig. \ref{gam_sum})
can be used to search for the C-violating decay
 $\phi\to\rho\gamma$, $\rho \to \pi^+\pi^-$. 
The cross section vs. energy for the events with photons in the 100-300 MeV
range is presented in Fig. \ref{rho-sech}. 
%
The final state in the C-violating  $\phi\to\rho\gamma$ decay has the same 
quantum numbers as in the initial bremsstrahlung process which 
dominates in the cross section and
to extract a
possible signal the interference of these two processes is assumed: \\

 $\sigma(s) = 
 \sigma^{in}_{br}(s)\cdot 
 | 1 - e^{i\psi}\cdot\sqrt
 {\frac{\sigma (\phi\to\rho\gamma)} {\sigma^{in}_{br}(s)}  } \cdot
 \frac{ m_{\phi}\Gamma_{\phi}} {\Delta_{\phi}} |^{2}$, \\

\noindent
where  $\sigma^{in}_{br}(s) = \sigma_{0} \cdot m^2_{\phi}/s$.
The comparison of 
the obtained cross section with the theoretical calculation \cite{acha97}
gives the ratio  $\sigma_{0}^{exp}/\sigma_{0}^{th}$=1.04$\pm$0.03. 
As a result of the fit,
 $Br(\phi\rightarrow\rho\gamma)=\sigma(\phi\to\rho\gamma)/\sigma_{tot}^{\phi} 
= (0.3\pm0.5)\times10^{-5}$ and $\psi=-0.9\pm 1.0$ 
radians have been obtained. Taking into account the 80\% detection
efficiency, the corresponding upper limit is:
  $Br(\phi\rightarrow\rho\gamma)< 1.2\times10^{-5}$ at 90\% C.L.

This result
should be compared to the previous measurements which gave upper limits
  $7\times10^{-4}$~\cite{CMDppg} and $2\times10^{-2}$~\cite{rhog}. 
\section*{ \boldmath Search for  $\eta\to\pi^+\pi^-$ Decay}
\hspace*{\parindent}
The selected  $\pi^+\pi^-\gamma$ 
events can be used to search for the
P- and CP- violating decay  $\eta\rightarrow\pi^+\pi^-$, where the 
 $\eta$ comes from
the radiative  $\phi\to\eta\gamma$ decay. 
From 19.7 millions of  $\phi$ decays used for the $\pi^+\pi^-\gamma$ channel
analysis 
one could expect about 248,000 events  
which decayed via the  $\eta\gamma$ channel.
These P-, CP-violating decays should  be observed as   
peaks  in the invariant mass of two pions at  $m_{\pi\pi}=m_{\eta}$.
%
\begin {figure}[tbh]
\epsfig{figure= 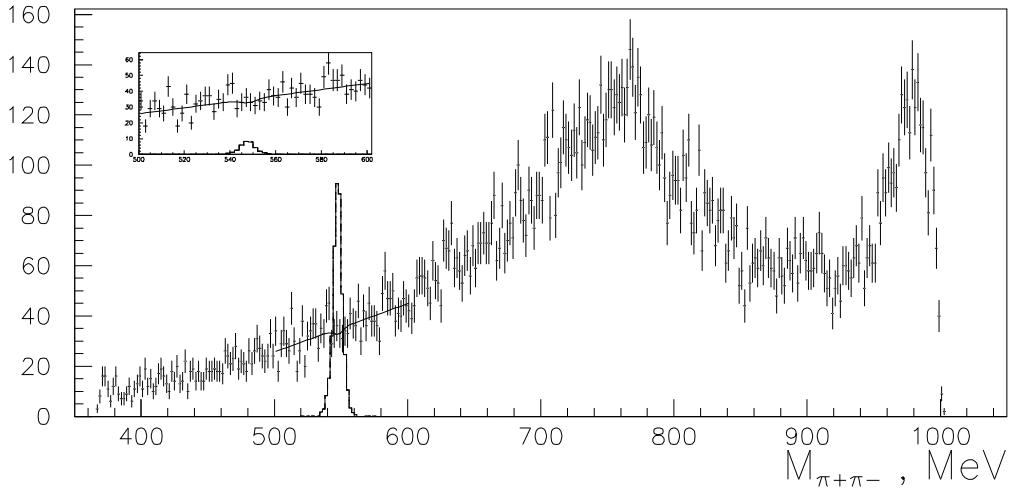, width=1.0\textwidth}
\caption{
Search for  $\eta\to\pi^+\pi^-$ decay. 
The histogram in the box shows a simulated possible signal at 90\%
CL.
} 
\label{etapipi}
\end{figure} 
Figure~\ref{etapipi} represents experimental distributions of
  $\pi^+\pi^-$ 
masses from selected $\pi^+\pi^-\gamma$ events.
The line corresponds to a fit with a linear function and gaussian
 distribution representing a possible signal.
It was found that the signal does not exceed  10 events at 90\% CL. 
The histogram in the box shows a simulated signal from
$\eta\rightarrow\pi^+\pi^-$  decays at 90\% CL. 
The detection efficiency found by simulation
was 0.124.
 The following result has been obtained:

Br($\eta\rightarrow\pi^{+} \pi^{-}) < 3.3\times 10^ {-4}$.\\
which should be compared to the best previous limit of 
  $9\times10^{-4}$ ~\cite{CMDppg}.

\section*{ \boldmath Conclusions}
\hspace*{\parindent}
Using 13.1 pb$^{-1}$ of data collected around
the   $\phi$ meson (about 20 millions of the $\phi$ decays)  
 $e^+e^-\to\pi^+\pi^-\gamma$ and  
 $e^+e^-\to\mu^+\mu^-\gamma$ events were selected. For the first time the
decay  $\phi\to\pi^+\pi^-\gamma$ has been observed in the
20-120 MeV photon energy range. The fit assuming that the $\phi$ 
contributes only to the photon propagator (no direct $\phi$
decay) gave the branching ratio:

 $Br(\phi\to\pi^+\pi^-\gamma)=(0.41\pm0.12\pm0.04)\times 10^{-4}$.\\
This value is nine times higher than the expected 
and points to the presence of the hadronic 
decay of the  $\phi$ into this final state. The analysis of the photon 
spectra shows the presence of a resonance in the $\pi^+\pi^-$ system
with a mass of about 980 MeV.
The obtained branching ratio can be affected by the complicated
interference of the hadronic $\phi$ decay with the bremsstrahlung process. 
 The effect of interference  was demonstrated by fitting 
photon spectra assuming the  
$\phi\to f_{0}(980)\gamma$ decay and the following results have been obtained:

 $Br(\phi\rightarrow f_{0}(980)\gamma)=(1.93\pm0.46\pm0.50)\times10^{-4}$;

 $m_{f_{0}}=976\pm5\pm6$ MeV;

 $\Psi = 1.55\pm0.22$ radians,\\
where $\Psi$ is the relative phase of the amplitudes of the $\phi$ decay 
and the bremsstrahlung process
taking into account the final state $\pi-\pi$ interaction.

According to the model~\cite{AchIvan,acha97,acha98}, the obtained 
branching ratio can only be explained if $f_{0}(980)$ is a four quark state.

For muons the value

  $Br(\phi\to\mu^+\mu^-\gamma)=(1.43\pm0.45\pm0.14)\times 10^{-5}$\\
has been obtained for the photon energies  $E_{\gamma}>20 MeV$ 
consistent with the theoretical expectations.

For the C-violating decay of  $\phi\to\rho\gamma$ and P-, CP-violating 
decay of  $\eta\rightarrow\pi^+\pi^-$  
the following upper limits at 90\% CL have been obtained:

  $Br(\phi\to\rho\gamma)~<~1.2\times10^{-5}$,

  $Br(\eta\rightarrow\pi^+\pi^-)~<~3.3\times10^{-4}$.\\
These results are the most stringent upper limits  at the moment~\cite{pdg}.

Analysis of the $f_0(980)\gamma$ intermediate state is continued in the 
following paper devoted to the studies of the $\pi^0\pi^0\gamma$ and
 $\eta\pi^0\gamma$ final states \cite{follow}.
\subsection*{Acknowledgements}
\hspace*{\parindent}
The authors are grateful to N.N.Achasov, V.P.Druzhinin, 
V.V.Gubin, V.N.Ivanchenko and A.I.Mil\-stein for useful
discussions and help with the data interpretation.
\end{document}